\documentclass[a4paper,12pt]{article}
\usepackage[pctex32]{graphics}

\begin{document}
\newcommand{\cl}{\centerline}
\renewcommand{\theequation}{\arabic{equation}}
\newcommand{\beq}{\begin{equation}}
\newcommand{\eeq}{\end{equation}}
\newcommand{\bea}{\begin{eqnarray}}
\newcommand{\eea}{\end{eqnarray}}
\newcommand{\nn}{\nonumber}
\newcommand\pa{\partial}
\newcommand\un{\underline}
\newcommand\ti{\tilde}
\newcommand\pr{\prime}
\begin{titlepage}
\setlength{\textwidth}{5.0in} \setlength{\textheight}{7.5in}
\setlength{\parskip}{0.0in} \setlength{\baselineskip}{18.2pt}

\begin{flushright}
{\tt SOGANG-MP 01/07}
\end{flushright}

\vspace*{0.05cm}

\begin{center}
{\large\bf Entropy of the Schwarzschild black hole to all orders in
the Planck length}
\end{center}

\begin{center}
{Yong-Wan Kim$^{1,a}$ and Young-Jai Park$^{2,b}$} \par
\end{center}

\begin{center}
{$^{1}$National Creative Research Initiative Center for Controlling
Optical Chaos, Pai-Chai University, Daejeon 302-735, Korea}\par
\vskip 0.2cm {$^{2}$Department of Physics and Mathematical Physics
Group,}\par {Sogang University, Seoul 121-742, Korea}\par
\end{center}
\vskip 0.05cm
\begin{center}
(\today)
\end{center}
\vskip 0.05cm
\begin{center}
{\bf ABSTRACT}
\end{center}
\begin{quotation}

Considering corrections to all orders in the Planck length on the
quantum state density from a generalized uncertainty principle
(GUP), we calculate the statistical entropy of the scalar field on
the background of the Schwarzschild black hole without any cutoff.
We obtain the entropy of the massive scalar field proportional to
the horizon area.

\vskip 0.1cm


\vskip 0.05cm

\noindent
$^a$ywkim@pcu.ac.kr \\
\noindent $^b$yjpark@sogang.ac.kr
\noindent

\end{quotation}
\end{titlepage}

\newpage

\section{Introduction}

Three decades ago, Bekenstein had suggested that the entropy of a
black hole is proportional to the area of the horizon through the
thermodynamic analogy \cite{bek}. Subsequently, Hawking showed that
the entropy of the Schwarzschild black hole satisfies exactly the
area law by means of Hawking radiation based on the quantum field
theory \cite{haw}. After their works, 't Hooft investigated the
statistical properties of a scalar field outside the horizon of the
Schwarzschild black hole by using the brick wall method with the
Heisenberg uncertainty principle (HUP) \cite{tho}. However, although
he obtained the entropy proportional to the horizon area, an
unnatural brick wall cutoff was introduced to remove the ultraviolet
divergence near the horizon \cite{gm,kkps,acb,kop,med,lz}. After
these works, many efforts \cite{gup,gup1} have been devoted to the
generalized uncertainty relations, which lead to the minimal length
as a natural ultraviolet cutoff \cite{snyder}, and its consequences,
especially the effect on the density of states.

Recently, in Refs. \cite{li,liu,kkp}, the authors calculated the
entropy of black holes to leading order in the Planck length by
using the newly modified equation of the density of states motivated
by the generalized uncertainty principle (GUP) \cite{gup}, which
drastically solves the ultraviolet divergences of the just vicinity
near the horizon without a cutoff. Moreover, Nouicer has investigated
the GUP effects to all orders in the Planck length on black hole
thermodynamics \cite{kn0} by arguing that the GUP up to leading
order correction in the Planck length is not enough because the wave
vector $k$ does not satisfy the asymptotic property in the modified
dispersion relation \cite{Sabine}. Very recently, he has extended
the calculation of entropy to all orders in the Planck length
\cite{kn} for the Randall-Sundrum brane case \cite{rs}.

On the other hand,  Yoon et. al. have very recently pointed out that
since the minimal length $\sqrt{\lambda}$ is actually related to the
brick wall cutoff $\epsilon$, the entropy integral about $r$ in the
range of the near horizon should be carefully treated for a
convergent entropy \cite{yoon}.

In this paper, we calculate the statistical entropy of a scalar
field on the Schwarzschild black hole background to all orders in
the Planck length by carefully considering the entropy integral
about $r$ in the range $(r_H, r_H + \epsilon)$ near the horizon. By
using the novel equation of the density of states \cite{kn0,kn}
motivated by the GUP in the quantum gravity, we calculate the
quantum entropy of a massive scalar field on the Schwarzschild black
hole background. As a result, we obtain the desired
Bekenstein-Hawking entropy without any artificial cutoff and little
mass approximation satisfying the asymptotic property of the wave
vector $k$ in the modified dispersion relation.

\section{All order corrections of GUP}

Now, it is well-known that the deformed Heisenberg algebra
\cite{gup} leads to the GUP showing the existence of the minimal
length. In this section we briefly recapitulate this approach and
exploit the recently obtained results
\cite{kn0,kn,Moffat,spallucci,nouicer}. Indeed, it has been shown
that the Feynman propagator displays an exponential ultraviolet
cutoff of the form of $\exp \left( -\lambda p^2\right) $, where the
parameter $\sqrt{\lambda}$ actually plays a role of the minimal
length as shown later. Recently, this framework has been further
applied to the black hole evaporation process \cite{nicolini,mkp}.
On the other hand, the quantum gravity phenomenology has been
tackled with effective models based on the GUPs and/or modified
dispersion relations \cite{Cam} containing the minimal length as a
natural ultraviolet cutoff \cite{Sabine}. Moreover, the essence of
the untraviolet finiteness of the Feynman propagator can be also
captured by a nonlinear relation $p=f(k)$, where $p$ and $k$ are the
momentum and the wave vector of a particle, respectively,
generalizing the commutator between the commutating operators
$\hat{x}$ and $\hat{p}$ to
\begin{equation}\label{xp}
[\hat{x}, \hat{p}] = i \frac{\partial p}{\partial k}~\Leftrightarrow
\Delta x \Delta p \geq \frac{1}{2} \left|~ \left< \frac{\partial
p}{\partial k} \right> ~\right|
\end{equation}
at the quantum mechanical level \cite{Sabine}. Then, the usual
momentum measure $\prod^n_{i=1} dp^i$ is deformed to
\begin{equation}\label{pm}
\prod^n_{j=1} dp^j \prod^n_{i=1} \frac{\partial k^i}{\partial p^j}.
\end{equation}
For simplicity, in the following, let us restrict ourselves to the
isotropic case in one space-like dimension. According to the Refs.
\cite{spallucci,nouicer}, we have
\begin{equation}
\frac{\partial p}{\partial k}=~e^{\lambda p^{2}}, \label{measure}
\end{equation}%
where $\lambda$ is a dimensionless constant of order one in the
Planck length units. Now, let us consider the following
representation of the position and momentum operators
\begin{eqnarray}
&& X \equiv i~e^{\lambda P^{2}} {\partial _{p}},\nonumber\\
&& P \equiv p.  \label{xp}
\end{eqnarray}
Then, these operators satisfy the deformed algebra as follows
\begin{equation}
\left[ X,P\right] =i~e^{\lambda P^{2}}, \label{comrel}
\end{equation}
which leads to the generalized uncertainty relation as
\begin{equation}
\Delta X \Delta P \geq \frac{1}{2}\left\langle e^{\lambda P^{2}}
\right\rangle. \label{xp1}
\end{equation}
Next, in order to investigate the quantum implication of this
deformed algebra, let us solve the above relation (\ref{xp1}) for
$\Delta P$ that is satisfied with the equality. Since $\left\langle
P^{2n}\right\rangle \geq \left\langle P^{2}\right\rangle ^{n}$ and
$\left( \Delta P\right)^{2}=\left\langle P^{2}\right\rangle
-\left\langle P\right\rangle^{2}$, the generalized uncertainty
relation can be written as
\begin{equation}
\label{gupr} \Delta X \Delta P =\frac{1}{2} e^{\lambda \left(
\left( \Delta P\right)^{2}+\left\langle P\right\rangle
^{2}\right)}.
\end{equation}
Taking the square of this expression and using the definition of the
multi-valued Lambert function \cite{Lambert} (see Fig.1), we obtain
\begin{equation}
W\left( \xi\right) e^{W\left( \xi\right) }=\xi,  \label{lam}
\end{equation}%
where we have set $W(\xi)=-2\lambda \left(\Delta P\right)^{2}$ and
$\xi=-\frac{\lambda}{{2\left( \Delta X\right) ^{2}}} e^{2\lambda
\left\langle P\right\rangle ^{2}}$.

\begin{figure}[t!]
   \centering
   \includegraphics{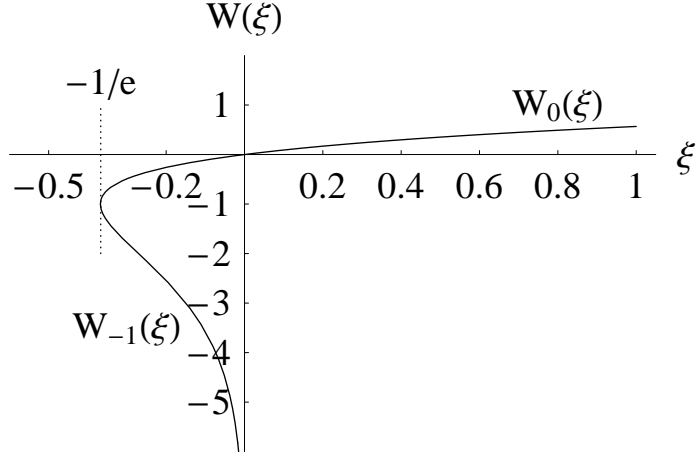}
\caption{Solutions of the Lambert function, $W\left( \xi\right)
e^{W\left( \xi\right) }=\xi$. When $\xi \geq 0$, the Lambert
function has only one real solution, $W_{0}(\xi)$. For $-1/e \leq
\xi < 0$, it has two real branches $W_{0}(\xi)$ and $W_{-1}(\xi)$
with a branch point $-1/e$. For $ -\infty < \xi < -1/e$, it has no
real solution.} \label{fig1}
\end{figure}

On the other hand, by using this function the momentum uncertainty
is given by
\begin{equation}
\Delta P =\frac{e^{\lambda \langle P \rangle^2 }}{2\Delta X}
e^{-\frac{1}{2}W(\xi)}. \label{argu}
\end{equation}
In order to have a real solution for $\Delta P$, the argument of the
Lambert function is required to satisfy $\xi \geq -1/e$, which leads
to the following condition
\begin{equation}
\frac{\lambda}{2(\Delta X)^2} e^{2\lambda \langle P \rangle^2 } \leq
\frac{1}{e}.
\end{equation}
This gives naturally the position uncertainty as
\begin{equation}
\Delta X \geq \sqrt{\frac{e\lambda}{2}}~e^{\lambda \langle P
\rangle^2} \equiv \Delta X _{\min},
\end{equation}
where $\Delta X_{\min }$ is a minimal uncertainty in position.
Moreover, this minimal length, which is intrinsically derived for
physical states with $\left\langle P\right\rangle =0$, is given by
\begin{equation}
\Delta X^A_{0}=\sqrt{\frac{e\lambda}{2}}. \label{min}
\end{equation}
This is the absolutely smallest uncertainty in position. In fact,
this minimal length effectively plays a role of the brick wall
cutoff giving the thickness of the thin-layer near the horizon
\cite{li,liu,kkp}. Furthermore, the momentum uncertainty with
$\left\langle P\right\rangle =0$ is easily read from Eq.
(\ref{argu}) with $W(\xi)=-2\lambda \left(\Delta P\right)^{2}$ as
\begin{equation}
\Delta P =\frac{1}{2\Delta X} e^{\lambda (\Delta P)^2}.
\label{allgup}
\end{equation}%
A series expansion of Eq. (\ref{allgup}) naturally includes the
well-known form of the GUP up to the leading order correction in the
Planck length units \cite{kkp} as follows
\begin{equation}
\label{gupL} \Delta X \Delta P \approx \frac{1}{2} \left[ 1+ \lambda
\left( \Delta P\right)^{2} + {\cal O}\left( \left( \Delta
P\right)^{4} \right) \right].
\end{equation}
Then, the minimal length up to the leading order is given by
\begin{equation}
\Delta X^L_0=\sqrt{\lambda}~<~\Delta X^A_{0} .
\end{equation}
However, only this leading order correction of the GUP does not
satisfy the property that the wave vector $k$ asymptotically reaches
the cutoff in large energy region as recently reported in Ref.
\cite{Sabine}.

In the following sections we use the form of the GUP given by Eq.
(\ref{allgup}) with the corresponding minimal length (\ref{min}) to
calculate the entropy of a scalar field on the Schwarzschild black
hole background. In this paper, we take the units
$G=\hbar=c=k_{B}\equiv 1$.

\section{Scalar field on the Schwarzschild black hole Background}

Let us consider the 4-dimensional Schwarzschild black hole solution
as
\begin{equation}
  ds^2 = - \left(1-\frac{2M}{r}\right) dt^2
        + \left(1-\frac{2M}{r}\right)^{-1} dr^2
        + r^2 d\Omega^2_{(2)},
 \label{d4ss}
\end{equation}
where $d\Omega^2_{(2)}$ is a metric of the unit 2-sphere. In this
background, let us first consider a scalar field with mass $\mu$,
which satisfies the Klein-Gordon equation given by
\begin{equation}
  \label{wveqn}
  (\nabla^2 - \mu^2)\Phi = 0.
\end{equation}
It can be rewritten as
\begin{equation}
  \label{wveqn2}
  -\frac{1}{f} \partial_{t}^2 \Phi + \frac{1}{r^2}
  \partial_{r}\left(r^2 f \partial_{r} \Phi\right) + \frac{1}{r^2 {\rm
  sin}\theta } \partial_{\theta} ({\rm sin} \theta \partial_{\theta}
  \Phi) + \frac{1}{r^2 {\rm sin}^2 \theta} \partial_{\phi}^2 \Phi
  - \mu^2 \Phi = 0
\end{equation}
with $f = 1- \frac{2M}{r}$. Substituting the wave function
$\Phi(t,r,\theta, \phi) = e^{-i\omega t}\psi(r, {\theta}, \phi)$, we
find that the Klein-Gordon equation becomes
\begin{equation}
\label{rtheta0} \partial_{r}^2 \psi + \left( \frac{1}{f}
\partial_{r} f + \frac{2}{r}\right)
\partial_{r} \psi + \frac{1}{f}
\left({\frac{1}{r^2}}\left[\partial^2_\theta + {\rm cot}\theta
\partial_\theta + {\frac{1}{{\rm sin}^{2}\theta}}\partial^2_\phi \right] +
\frac{\omega^{2}}{f} - \mu^{2} \right)\psi = 0.
\end{equation}
By using the Wenzel-Kramers-Brillouin approximation \cite{tho} with
$\psi \sim exp[iS(r,\theta,\phi)]$, we have
\begin{equation}
\label{wkb} {p_{r}}^{2} = \frac{1}{f}\left(\frac{\omega^{2}}{f} -
\mu^{2} - \frac{p^2_\theta}{r^2} - \frac{p_{\phi}^2}{r^2 {\rm sin}^2
\theta} \right),
\end{equation}
where
\begin{equation}
\label{mom} p_{r} = \frac{\partial S}{\partial r},~
 p_{\theta} = \frac{\partial S}{\partial \theta},
 ~p_{\phi} = \frac{\partial S}{\partial \phi}.
\end{equation}
Furthermore, we also obtain the square module momentum as follows
\begin{equation}
\label{smom} p^{2} = p_{i}p^{i} = g^{rr}{p_{r}}^{2} + g^{\theta
\theta}{p_{\theta}}^{2}+ g^{\phi\phi}{p_{\phi}}^{2} =
\frac{\omega^{2}}{f} - \mu^{2}.
\end{equation}
Then, the volume in the momentum phase space is given by
\begin{eqnarray}
V_{p}(r,\theta)&=& \int dp_{r}dp_{\theta}dp_{\phi} \nonumber \\
&=& \frac{4\pi}{3} \sqrt{ \frac{1}{f}(\frac{\omega^2}{f}-\mu^2)}
\cdot \sqrt{r^2(\frac{\omega^2}{f}-\mu^2)} \cdot \sqrt{{r^2 {\rm
sin}^2 \theta}(\frac{\omega^{2}}{f} - \mu^{2})} \nonumber
\\&=&\frac{4\pi}{3} \frac{r^2 {\rm sin}\theta}{\sqrt{f}}
\left(\frac{\omega^2}{f}-\mu^2 \right)^{\frac{3}{2}}
\end{eqnarray}
with the condition $\omega\geq\mu\sqrt{f}$.

\section{Entropy to all orders in the Planck Length}
Now, let us calculate the statistical entropy of the scalar field on
the Schwarzschild black hole background to all orders in the Planck
length units. When the gravity is turned on, the number of quantum
states in a volume element in phase cell space based on the GUP in
the 3+1 dimensions is given by
\begin{equation}
\label{dn} dn_A = \frac{d^3 x d^3 p}{(2\pi)^3}e^{-\lambda p^2} ,
\end{equation}
where $p^2 = p^{i}p_{i}~(i = r, \theta, \phi)$ and one quantum state
corresponding to a cell of volume is changed from $(2\pi)^3 $ into
$(2\pi)^3 e^{\lambda p^2}$ in the phase space \cite{li,liu,kkp}.

From the Eqs. (\ref{smom}) and (\ref{dn}), the number of quantum
states related to the radial mode with energy less than $\omega$ is
given by
\begin{eqnarray}
\label{Tnqs} n_A(\omega) &=& \frac{1}{(2\pi)^3} \int dr d\theta
d\phi
dp_{r} dp_{\theta}dp_{\phi} e^{- \lambda p^2} \nonumber \\
&=&\frac{1}{(2\pi)^3} \int dr d\theta d\phi V_{p}(r,\theta) e^{- \lambda (\frac{{\omega}^2}{f}- \mu^{2})} \nonumber   \\
&=& \frac{2}{3\pi}\int_{r_{H}} dr \frac{r^2}{\sqrt{f}}
\left(\frac{{\omega}^2}{f}- \mu^{2}\right)^{\frac{3}{2}} e^{-\lambda
(\frac{{\omega}^2}{f}- \mu^{2})}.
\end{eqnarray}
It is interesting to note that $n_A(\omega)$ is convergent at the
horizon without any artificial cutoff due to the existence of the
suppressing exponential $\lambda$-term induced from the generalized
uncertainty principle.

For the bosonic case, the free energy at inverse temperature $\beta$
is given by
\begin{equation}
\label{def} e^{-\beta F} = \prod_K
                \left[ 1 - e^{-\beta \omega_K} \right]^{-1}~,
\end{equation}
where $K$ represents the set of quantum numbers. Then, by using Eq.
(\ref{Tnqs}), we are able to obtain the free energy as
\begin{eqnarray}
\label{TfreeE}
 F_{A}&=& \frac{1}{\beta}\sum_K \ln \left[ 1 - e^{-\beta \omega_K} \right]
   ~\approx ~\frac{1}{\beta} \int dn_A(\omega) ~\ln
            \left[ 1 - e^{-\beta \omega} \right]  \nonumber   \\
 &=& - \int^{\infty}_{\mu\sqrt{f}} d\omega
 \frac{n_A(\omega)}{e^{\beta\omega} -1} \nonumber  \\
   &=& - \frac{2}{3\pi} \int_{r_{H}} dr \frac{r^2}{\sqrt{f}}
   \int^{\infty}_{\mu\sqrt{f}} d\omega~
    \frac{\left(\frac{{\omega}^2}{f}
   - \mu^{2}\right)^{\frac{3}{2}}}{(e^{\beta \omega} -1)}e^{- \lambda (\frac{{\omega}^2}{f}- \mu^{2})}.
\end{eqnarray}
Here, we have taken the continuum limit in the first line and
integrated it by parts in the second line. In the last line, since
$f \rightarrow 0$ near the event horizon, {\it i.e.}, in the range
of $(r_H, r_H +\epsilon)$, $\frac{{\omega}^2}{f}
   - \mu^{2}$ becomes $\frac{{\omega}^2}{f}$.
Therefore, although we do not require the little mass approximation,
the free energy can be rewritten as \\
\beq \label{TfreeEf0}
 F_{A} = - \frac{2}{3\pi} \int^{r_{H}+\epsilon}_{r_{H}} dr
\frac{r^2}{f^2} \int^{\infty}_{0} d\omega
\frac{{\omega}^3}{(e^{\beta \omega} -1)}
 e^{- \lambda {\frac{{\omega}^2}{f}}}. \eeq

On the other hand, we are also interested in the contribution from
just the vicinity near the horizon in the range $(r_H, r_H +
\epsilon)$, where $\epsilon$ is related to a proper distance of
order of the minimal length (\ref{min}) as follows
\begin{eqnarray}
\label{invariant} \sqrt{\frac{e\lambda}{2}} =
             \int_{r_H}^{r_H+\epsilon} \frac{dr}{\sqrt{f(r)}}
                \approx \int_{r_H}^{r_H +\epsilon}
                          \frac{dr}{\sqrt{2\kappa(r-r_{H})}}
                 = \sqrt{\frac{2\epsilon}{\kappa}},
\end{eqnarray}
where the expansion of $f(r)$ near the horizon is given by
\begin{equation}
\label{Tay} f(r) \approx f(r_{H}) +
(r-r_{H})\left(\frac{df}{dr}\right)|_{\beta =\beta_{H}} + {\cal
O}\left( (r-r_{H})^2 \right).
\end{equation}
Here, $\kappa$ is the surface gravity at the horizon of the black
hole, and it is identified as $\kappa
=\frac{1}{2}(\frac{df}{dr})|_{\beta =\beta_{H}} = 2\pi
{\beta_H}^{-1} = 1/(2r_{H})$.

Before calculating the entropy, let us mention that Yoon et. al.
have recently suggested that since the minimal length
$\sqrt{\frac{e\lambda}{2}}$ in Eq. (\ref{invariant}) is related to
the brick wall cutoff $\epsilon$, the entropy integral about $r$ in
the range near the horizon should be carefully treated for
obtaining a convergent entropy \cite{yoon}. In particular, although
the term $(e^{\beta \omega}-1)$ in Eq. (\ref{TfreeEf0}) with $x=
\sqrt{\frac{\lambda}{f}}\omega$ was expanded in the previous works
giving $\beta\sqrt{\frac{f}{\lambda}}x$, one may not simply expand
up to the first order since $0\leq \frac{f}{\lambda} =
\frac{2\kappa(r-r_H)}{\lambda} \leq \frac{2 \kappa
\epsilon}{\lambda} = \kappa^2$ near the horizon.

Now, let us carefully consider the integral about $r$ near the
horizon by extracting out the $\epsilon$-factor through Taylor's
expansion (\ref{Tay}) of $f(r)$. Then, from $F_A$ in Eq.
(\ref{TfreeEf0}) the entropy can be obtained as

\begin{eqnarray}
\label{Aentropy0}
S_{A} &=& \beta^2 \frac{\partial F_A}{\partial \beta}\mid_{\beta=\beta_H} \nonumber \\
&=& \frac{\beta^2_H}{6\pi} \int^{\infty}_{0} d\omega
\frac{{\omega}^4 }{\sinh^2(\frac{\beta_H}{2}\omega)}\int^{r_{H} +
\epsilon}_{r_{H}} dr
\frac{r^2}{f^2} e^{- \lambda {\frac{{\omega}^2}{f}}} \nonumber \\
&=& \frac{\beta^2_H}{6\pi \lambda^2\sqrt{\lambda}} \int^{\infty}_{0}
dx \frac{{x}^4
}{\sinh^2(\frac{\beta_H}{2\sqrt{\lambda}}x)}\Lambda_{A}(x,\epsilon),
\end{eqnarray}
where $x\equiv \sqrt{\lambda}\omega$ and the integral about $r$ in
the range of the near horizon is given by
\begin{eqnarray}
\label{RfreeEf}
 \Lambda_{A}(x,\epsilon) & = & \int^{r_{H} + \epsilon}_{r_{H}} dr~
\frac{r^2}{f^2}~e^{- {\frac{{x}^2}{f}}} = \int^{r_{H} +
\epsilon}_{r_{H}} dr~r^2~\frac{1}{f^2 \left[ \Sigma^\infty_{n=0}\frac{1}{n!} \left(\frac{{x}^2}{f}\right)^n \right]} \nonumber \\
 & = & \frac{6}{x^6}\int^{r_{H} + \epsilon}_{r_{H}} dr \left[r^2_H + 2r_H (r-r_H) + {\cal O}\left((r-r_H)^2 \right) \right] \nonumber \\
 &  & ~~~~~~~~~~~~\times\left(\frac{2\kappa(r-r_H) +  {\cal O}\left((r-r_H)^2 \right)}
    {1 + \frac{2\kappa}{3x^2}(r-r_H) +  {\cal O}\left((r-r_H)^2 \right)}\right) \nonumber \\
& \approx & \frac{12\kappa r^2_{H}}{x^6} \int^{r_{H} +
\epsilon}_{r_{H}}
dr \left[(r-r_H) + {\cal O}\left((r-r_H)^2 \right) \right] \nonumber \\
&\approx& \frac{6 \kappa r^2_{H}\epsilon^2}{x^6}.
\end{eqnarray}
Then, the entropy is reduced as
\begin{eqnarray}
\label{Aentropy1} S_{A} &\approx& \frac{\beta^2_H}{6\pi
\lambda^2\sqrt{\lambda}} \int^{\infty}_{0} dx \frac{{x}^4
}{\sinh^2(\frac{\beta_H}{2\sqrt{\lambda}}x)}~\frac{6 \kappa
r^2_{H}\epsilon^2}{x^6} \nonumber \\
&=& \frac{\pi^2 e^2 r^2_{H}}{4\lambda} \int^{\infty}_{0} dy
\frac{1}{{y}^2 \sinh^{2}y},
\end{eqnarray}
where $y \equiv \frac{\beta_H}{2\sqrt{\lambda}}x$, $\beta_H \kappa =
2\pi$, and $\epsilon = \lambda\kappa/4$.

Therefore, when $r\rightarrow r_{H}$, we finally get the desired
entropy of the massive scalar field on the Schwarzschild black
background as follows
\begin{eqnarray}
\label{finalS} S_{A} &\approx& \frac{\pi^2 e^2
r^2_{H}}{4\lambda}~\frac{2\zeta(3)}{\pi^2}
= \frac{e^2 \zeta(3)}{8\pi\lambda}(4\pi r^2_{H})\nonumber \\
&=& ~\frac{1}{4}~\frac{e^2 \zeta(3)}{2\pi\lambda}A,
\end{eqnarray}
where $A= 4 \pi r_{H}^2$ and $\zeta(3)= \Sigma^{\infty}_{n=1}
(1/n^3)~\approx 1.202$. Moreover, if we assume the minimal length
parameter $\lambda$ in the Planck length units as $\frac{e^2
\zeta(3)}{2\pi}$, then the entropy can be rewritten by the desired
area law as $S_{A} = \frac{1}{4}~A$. Note that there is no divergence
within the just vicinity near the horizon due to the effect of the
generalized uncertainty relation on the quantum states.

On the other hand, in order to compare the result (\ref{finalS})
with those of the usual  approximation approach
\cite{li,liu,kkp,kn}, let us calculate the entropy in the usual
coarse-grained approximation. In terms of the variable $x=\omega
\sqrt{\lambda /f}$ \ and the fact that $e^{\beta \omega }-1=e^{\beta
\sqrt{\frac{f}{\lambda }}x}-1\approx \beta
\sqrt{\frac{f}{\lambda}}x$ for $f\rightarrow 0$, we have \\
\begin{eqnarray} \label{TfreeEf1}
 F^0_{A} & = & - \frac{1}{\beta}\frac{2}{3\pi (\lambda)^{3/2}} \int^{r_{H}+\epsilon}_{r_{H}} dr
              \frac{r^2}{\sqrt{f}} \int^{\infty}_{0} dx x^2 e^{-x^2} \nonumber \\
         & = & - \frac{1}{\beta}\frac{1}{6 \sqrt{\pi} (\lambda)^{3/2}}
                 \int^{r_{H}+\epsilon}_{r_{H}} dr \frac{r^2}{\sqrt{f}}.
\end{eqnarray}
Then, from the free energy (\ref{TfreeEf1}), the entropy to all
orders for the scalar field is given by
\begin{eqnarray}
\label{Aentropy0}
S^0_{A} &=& \beta^2 \frac{\partial F_T}{\partial \beta}\mid_{\beta=\beta_H} \nonumber \\
        &=& \frac{1}{6 \sqrt{\pi} \lambda^{3/2}}
                \int^{r_{H}+\epsilon}_{r_{H}} dr \frac{1}{\sqrt{f}}~r^2 \nonumber \\
       & \approx & \frac{1}{6 \sqrt{\pi} \lambda^{3/2}} \sqrt{e \lambda/2}
           ~r^2_H = \frac{\sqrt{e}}{24\sqrt{2} \pi^{3/2} \lambda }A.
\end{eqnarray}
This is smaller than $S_A$ in Eq.(\ref{finalS}), which was obtained
by using the rigorous approximation near the horizon. But, if we
assume the minimal length parameter $\lambda$ as $
\frac{\sqrt{e}}{6\sqrt{2} \pi^{3/2}}$, then the entropy can be also
rewritten by the desired Bekenstein-Hawking area law as $S^0_{A} =
\frac{1}{4}~A$.

Finally, it seems to be appropriate to comment on the entropy
(\ref{finalS}) to all orders in the Planck length comparing with the
entropy to the leading order, which can be also carefully treated
through the same expansion approach. In this case, the free energy
up to the leading order \cite{li} is given by
\begin{eqnarray} \label{freeEfL}
 F_{L} & \approx & - \frac{2}{3\pi} \int^{r_{H}+ \epsilon}_{r_{H}} dr
\frac{r^2}{f^2} \int^{\infty}_{0} d\omega
\frac{{\omega}^3}{(e^{\beta \omega} -1)
 \left(1 + \frac{\lambda}{f}\omega^2 \right)^3}
 \end{eqnarray}
instead of $F_{A}$ in Eq.(\ref{TfreeEf0}). Then, the entropy $S_{L}$
is given by
\begin{eqnarray}
\label{LfreeEnoM}
  S_L &=& \beta^2 \frac{\partial F_L}{\partial \beta}\mid_{\beta=\beta_H}  \nonumber \\
  &=& \frac{\beta^2_H}{6\pi}
   \int^{\infty}_{0} d\omega
   \frac{\omega^4}{\sinh^2(\frac{\beta_H}{2}\omega)}
 \int^{r_{H}+ \epsilon}_{r_{H}} dr \frac{r^2}{f^2
 \left(1 + \frac{\lambda}{f}\omega^2 \right)^3} \nonumber \\
 &=& \frac{\beta^2_H}{6\pi \lambda^2\sqrt{\lambda}}\int^{\infty}_{0} dx
   \frac{x^4}{\sinh^2(\frac{\beta_H}{2\sqrt{\lambda}}x)} \Lambda_{L}(x,\epsilon),
\end{eqnarray}
where $x= \sqrt{\lambda}\omega$ and $\Lambda_{L}(x,\epsilon)$ is
given by
\begin{eqnarray} \label{IL} \Lambda_{L}(x,\epsilon) = \frac{1}{x^6}~\int^{r_{H}+
\epsilon}_{r_{H}} dr~\frac{r^2 f}{\left(1 + \frac{f}{x^2}
\right)^3}.
\end{eqnarray}
Then, the integral (\ref{IL}) is obtained as
\begin{eqnarray}
\label{IL1} \Lambda_{L}(x,\epsilon) &=& \frac{r^2_H}{x^6}
\int^{r_{H}+ \epsilon}_{r_{H}} dr \frac{ 2\kappa (r - r_H) + {\cal
O}\left((r-r_H)^2\right) }{
 \left[1 + \frac{2\kappa}{x^2}(r - r_H) + {\cal O}\left( (r-r_H)^2 \right) \right]^3} \nonumber \\
 & = & \frac{2\kappa r^2_{H}}{x^6} \int^{r_{H} + \epsilon}_{r_{H}}
dr \left[(r-r_H) + {\cal O}\left((r-r_H)^2 \right) \right] \nonumber \\
&\approx& \frac{\kappa r^2_{H}\epsilon^2}{x^6}.
\end{eqnarray}
As a result, the entropy up to the leading order correction becomes
\begin{eqnarray}
\label{LfreeEnoM1}
  S_L &\approx& ~\frac{1}{4}~\frac{\zeta(3)}{3\pi\lambda}~A < S_A.
\end{eqnarray}

\section{Summary}

We have investigated the entropy (\ref{finalS}) to all orders in the
Planck length units through the rigorous Taylor expansion approach
comparing with the coarse-grained entropy (\ref{Aentropy0}), which
has been obtained through the usual approximation approach. Although
their values are different, we have obtained the desired
Bekenstein-Hawking entropy by properly adjusting the minimal length
parameter $\lambda$ for the both approximation approaches.

In summary, we have studied the massive scalar field on the
background of the Schwarzschild black hole by carefully counting the
number of quantum states in the just vicinity near the horizon,
based on the generalized uncertainty principle. As a result, we have
obtained the desired Bekenstein-Hawking entropy to all orders in the
Planck length units without any artificial cutoff and little mass
approximation satisfying the asymptotic property of the wave vector
$k$ in the modified dispersion relation.

\section*{Acknowledgments}

We would like to thank Prof. W. Kim and Dr. Yoon for useful
discussions. This work is supported by the Kosong Research Grant for
Mathematical Physics.

\end{document}